# Theoretical Analysis of Generalized Sagnac Effect in the Standard Synchronization

**Yang-Ho Choi**

**Department of Electrical and Electronic Engineering**
**Kangwon National University**
**Chunchon, Kangwon-do, 200-701, South Korea**

**Abstract:** The Sagnac effect has been shown in inertial frames as well as rotating frames. We solve the problem of the generalized Sagnac effect in the standard synchronization of clocks. The speed of a light beam that traverses an optical fiber loop is measured with respect to the proper time of the light detector, and is shown to be other than the constant *c*, though it appears to be *c* if measured by the time standard-synchronized. The fiber loop, which can have an arbitrary shape, is described by an infinite number of straight lines such that it can be handled by the general framework of Mansouri and Sexl (MS). For a complete analysis of the Sagnac effect, the motion of the laboratory should be taken into account. The MS framework is introduced to deal with its motion relative to a preferred reference frame. Though the one-way speed of light is other than *c*, its two-way speed is shown to be *c* with respect to the proper time. The theoretical analysis of the generalized Sagnac effect corresponds to the experimental results, and shows the usefulness of the standard synchronization. The introduction of the standard synchrony can make mathematical manipulation easy and can allow us to deal with relative motions between inertial frames without information on their velocities relative to the preferred frame.

**Keywords:** Generalized Sagnac effect, Standard synchronization, Speed of light, Test theory, General framework for transformation

**PACS numbers:** 03.30.+p, 02.10.Yn, 42.25.Bs

# 1. Introduction

The Sagnac effect is a phenomenon of interference by two light beams propagating along a closed loop in opposite directions. It had usually been recognized to occur in circular motion which accompanies acceleration, before observed in the experiments [1, 2] that involve uniform linear motion as well. The Sagnac effect may be still a conundrum to the theory of relativity which is based on the constancy of the speed of light. Even the speeds of the counter-propagating light beams in the rotating frame are not clearly known, though there are various explanations and analyses available in the literature [e.g., 3–5 and references therein]. As for the generalized Sagnac effect that involves linear motion as well as circular motion, on the contrary, few expositions or analyses can be found [6]. The generalized Sagnac effect may be a more puzzling problem to the theory of relativity.

Recently, presupposing a preferred reference system $S$ the space-time space of which is isotropic so that the speed of light is constant irrespective of its propagation direction, a coordinate transformation between $S$ and a rotating system [7], called the transformation under the constant light speed (TCL), has been presented. TCL not only holds the constancy of the two-way speed of light in the rotating system but also is consistent with the transformation for inertial systems, which is derived from the former in the limit to inertial motion. The generalized Sagnac effect can be analyzed through TCL. Though the analytic results via TCL correspond to the experimental results, the analysis has been made under the constraints that the laboratory frame is assumed to be identical with the isotropic system $S$ and the optical fiber loop which the counter-propagating light beams traverse has a simple shape composed of circular and linear paths only.

In this paper, the problem of the generalized Sagnac effect is, without the constraints, solved based on the test theory of Mansouri and Sexl (MS) [8]. The MS test theory also presumes a preferred reference frame $S$. The motion of the laboratory should be considered for a complete analysis of the Sagnac effect. We introduce the general framework to deal with its motion relative to $S$, not with the problem of clock synchronization. The standard synchronization is adopted within the framework so that the speed of light appears to be constant with respect to the time standard-synchronized, which will be referred to as adjusted time (AT). We analyze the speeds of light beams traveling around an optical fiber loop which rotates at a constant speed relative to the laboratory frame, measuring them with respect to the proper time (PT). The fiber loop can have an arbitrary shape, and is described by an infinite number of differential lines such that the MS framework can handle it, regarding each differential line as an inertial frame. The analysis results correspond with the results of the experiments including the cases where the motions of the light beams and the fiber loop are non-collinear.

## 2. General Framework for Transformation

Consider an inertial frame $S'$ that is in rectilinear motion at a normalized velocity $\boldsymbol{\beta} = \mathbf{v}/c$ relative to $S$ where the speed of light is a constant $c$ regardless of its propagation direction. The MS test theory [8] provides a general framework for the transformation between $S'$ and the isotropic frame $S$. The space-time coordinate vector of $S'$ is represented as $\boldsymbol{p}' = [\tau', \boldsymbol{x}'^T]^T$ where $\tau' = ict'$ with $i = (-1)^{1/2}$, $\boldsymbol{x}'$ is a spatial vector, and $T$ stands for the transpose. The coordinate vector of $S$ is similarly represented without primes. For a vector $\boldsymbol{q}$, we denote its normalized vector by $\hat{\boldsymbol{q}}$, i.e., $\hat{\boldsymbol{q}} = \boldsymbol{q} / \|\boldsymbol{q}\|$, and its magnitude by $q$, i.e., $q = \|\boldsymbol{q}\|$, where $\|\cdot\|$ designates the Euclidean norm. In the MS framework, the coordinates of $S$ are transformed into $S'$ as follows [8–10]:

$$\tau' = a\tau + i\boldsymbol{\varepsilon}\cdot\boldsymbol{x}' = g\tau + i\boldsymbol{\rho}^T\boldsymbol{x} \tag{1a}$$

$$\boldsymbol{x}' = ib\tau\boldsymbol{\beta} + b\hat{\boldsymbol{\beta}}(\hat{\boldsymbol{\beta}}^T\boldsymbol{x}) + d(\boldsymbol{x} - \hat{\boldsymbol{\beta}}(\hat{\boldsymbol{\beta}}^T\boldsymbol{x})) \tag{1b}$$

where

$$g = a - b\boldsymbol{\varepsilon}^T\boldsymbol{\beta} \tag{2}$$

$$\boldsymbol{\rho} = (b-d)(\boldsymbol{\varepsilon}^T\hat{\boldsymbol{\beta}})\hat{\boldsymbol{\beta}} + d\boldsymbol{\varepsilon} \tag{3}$$

and $\boldsymbol{\varepsilon}$ is determined by a synchronization scheme in $S'$. The transformation coefficients $a$ and $b$ are associated with time dilation and length contraction, respectively. The general transformation can be represented in matrix form as

$$\boldsymbol{p}' = \boldsymbol{T}_G\boldsymbol{p} \tag{4}$$

where $\boldsymbol{T}_G$ can be expressed as a partitioned matrix [9]:

$$\boldsymbol{T}_G = \begin{bmatrix} g & i\boldsymbol{\rho}^T \\ ib\boldsymbol{\beta} & \boldsymbol{M}(\boldsymbol{\beta}) \end{bmatrix} \tag{5}$$

where

$$\boldsymbol{M}(\boldsymbol{\beta}) = (b-d)\hat{\boldsymbol{\beta}}\hat{\boldsymbol{\beta}}^T + d\boldsymbol{I} \tag{6}$$

with $\boldsymbol{I}$ an identity matrix.

Given $\tau'$ and $\boldsymbol{x}'$, one can obtain $\tau$ and $\boldsymbol{x}$ from them by using the inverse of $\boldsymbol{T}_G$. From (1), $\tau$ and $\boldsymbol{x}$ are related to $\tau'$ and $\boldsymbol{x}'$ by [9, 10]

$$\tau == \frac{1}{a}(\tau' - i\boldsymbol{\varepsilon}^T\boldsymbol{x}') \tag{7a}$$

$$\boldsymbol{x} = \frac{1}{d}\boldsymbol{x}' + (\frac{1}{b} - \frac{1}{d})(\hat{\boldsymbol{\beta}}^T\boldsymbol{x}')\hat{\boldsymbol{\beta}} - \frac{1}{a}(i\tau' + \boldsymbol{\varepsilon}^T\boldsymbol{x}')\boldsymbol{\beta}\,. \tag{7b}$$

The transformation (7) is rewritten in matrix form as

$$\boldsymbol{p} = \boldsymbol{T}_G^{-1}\boldsymbol{p}' \tag{8}$$

where

$$\boldsymbol{T}_G^{-1} = \begin{bmatrix} a^{-1} & -ia^{-1}\boldsymbol{\varepsilon}^T \\ -ia^{-1}\boldsymbol{\beta} & \boldsymbol{M}'(\boldsymbol{\beta}) \end{bmatrix} \tag{9}$$

$$\boldsymbol{M}'(\boldsymbol{\beta}) = (\frac{1}{b}-\frac{1}{d})\hat{\boldsymbol{\beta}}\hat{\boldsymbol{\beta}}^T - \frac{1}{a}\boldsymbol{\beta}\boldsymbol{\varepsilon}^T + \frac{1}{d}\boldsymbol{I}\,. \tag{10}$$

Let us find the transformation between arbitrary inertial frames, say $S_i$ and $S_j$, by using the general transformation. An inertial frame $S_k$, $k = i, j$, is in uniform linear motion at a normalized velocity $\boldsymbol{\beta}_k$ relative to $S$ and its coordinate vector is denoted as $\boldsymbol{p}_{(k)} = [\tau_{(k)}, x_{(k)}, y_{(k)}, z_{(k)}]$. The coordinate vector of $S_k$ is related to that of $S$, according to (4), by $\boldsymbol{p}_{(k)} = \boldsymbol{T}_G(\boldsymbol{\beta}_k)\boldsymbol{p}$ where the dependence of the transformation on the velocity is explicitly expressed as two or more inertial frames are involved. The dependency of transformation coefficients on velocities such as $a(\boldsymbol{\beta}_k)$ is also expressed explicitly. The transformation between $\boldsymbol{p}_{(j)}$ and $\boldsymbol{p}_{(i)}$ is then written as [9]

$$\boldsymbol{p}_{(j)} = \boldsymbol{T}_G(\boldsymbol{\beta}_j, \boldsymbol{\beta}_i)\boldsymbol{p}_{(i)} \tag{11}$$

where

$$\boldsymbol{T}_G(\boldsymbol{\beta}_j, \boldsymbol{\beta}_i) = \boldsymbol{T}_G(\boldsymbol{\beta}_j)\boldsymbol{T}_G^{-1}(\boldsymbol{\beta}_i)\,. \tag{12}$$

It is straightforward to calculate $\boldsymbol{T}_G(\boldsymbol{\beta}_j, \boldsymbol{\beta}_i)$ by using (5), (9), and (12):

$$\boldsymbol{T}_G(\boldsymbol{\beta}_j, \boldsymbol{\beta}_i) = \begin{bmatrix} a_i^{-1}(g_j + \boldsymbol{\rho}_j^T\boldsymbol{\beta}_i) & -ig_j a_i^{-1}\boldsymbol{\varepsilon}_i^T + i\boldsymbol{\rho}_j^T\boldsymbol{M}'(\boldsymbol{\beta}_i) \\ ia_i^{-1}(b_j\boldsymbol{\beta}_j - \boldsymbol{M}(\boldsymbol{\beta}_j)\boldsymbol{\beta}_i) & a_i^{-1}b_j\boldsymbol{\beta}_j\boldsymbol{\varepsilon}_i^T + \boldsymbol{M}(\boldsymbol{\beta}_j)\boldsymbol{M}'(\boldsymbol{\beta}_i) \end{bmatrix} \tag{13}$$

where $g_k = a_k - b_k\boldsymbol{\varepsilon}_k^T\boldsymbol{\beta}_k$ and $\boldsymbol{\rho}_k = (b_k - d_k)(\boldsymbol{\varepsilon}_k^T\hat{\boldsymbol{\beta}}_k)\hat{\boldsymbol{\beta}}_k + d_k\boldsymbol{\varepsilon}_k$ with $a_k = a(\boldsymbol{\beta}_k)$, $b_k = b(\boldsymbol{\beta}_k)$, and $d_k = d(\boldsymbol{\beta}_k)$. Given $\boldsymbol{\beta}_i$ and $\boldsymbol{\beta}_j$, the velocity $\boldsymbol{\beta}_{ji}$ in $S_i$ of an object $O_j$ which is at rest in $S_j$ is given by [10]

$$\boldsymbol{\beta}_{ji} = \frac{1}{a_j\Gamma(\boldsymbol{\beta}_i, \boldsymbol{\beta}_j)}(b_i[\hat{\boldsymbol{\beta}}_i(\hat{\boldsymbol{\beta}}_i^T\boldsymbol{\beta}_j) - \boldsymbol{\beta}_i] + d_i[\boldsymbol{\beta}_j - \hat{\boldsymbol{\beta}}_i(\hat{\boldsymbol{\beta}}_i^T\boldsymbol{\beta}_j)]) \tag{14}$$

where

$$\Gamma(\boldsymbol{\beta}_i, \boldsymbol{\beta}_j) = \frac{a_i}{a_j} + \frac{b_i}{a_j}(\boldsymbol{\varepsilon}_i^T[\hat{\boldsymbol{\beta}}_i(\hat{\boldsymbol{\beta}}_i^T\boldsymbol{\beta}_j) - \boldsymbol{\beta}_i]) + \frac{d_i}{a_j}(\boldsymbol{\varepsilon}_i^T[\boldsymbol{\beta}_j - \hat{\boldsymbol{\beta}}_i(\hat{\boldsymbol{\beta}}_i^T\boldsymbol{\beta}_j)])\,. \tag{15}$$

It is worth noting that the direction of $\boldsymbol{\beta}_{ji}$ is independent of the synchronization vector, though its magnitude is dependent. One can easily see from (14) that $\boldsymbol{\beta}_{ji}$ is reduced to $\boldsymbol{\beta}_j$ when $\boldsymbol{\beta}_i = \mathbf{0}$ so that $a_i = b_i = d_i = 1$ and $\boldsymbol{\varepsilon}_i = \mathbf{0}$.

## 3. Analysis of Generalized Sagnac Effect

In the experiments of the generalized Sagnac effect, two light beams which left an emitting source at the same time traverse an optical fiber loop in opposition directions. We investigate the speeds of the light beams and the difference between their arrival times, assuming that the Laboratory frame is isotropic in Subsection 3.1 and considering its motion relative to the isotropic preferred frame $S$ in Subsection 3.2. In [7], using TCL, the generalized Sagnac effect has been dealt with under the conditions described in Subsection 3.1.

### 3.1. Under the Assumption of Isotropic Laboratory Frame

We assume that the laboratory frame is isotopic so that the speed of light is $c$ in every direction, which allows us to denote it by $S$. First consider a simple shape of the optical fiber loop as shown in Fig. 1(a) (in Subsection 3.2, a loop of arbitrary shape is dealt with). The fiber loop rotates at a speed of $v$. The emitting source of light is placed at $P_0$ in Fig. 1(a). Two counter-propagating light beams travel around the loop and are received by a detector $O'$ located at $P_5$, which is the same place as $P_0$. We denote the co- and counter-rotating light beams by $b_+$ and $b_-$, respectively. The loop can be divided into two parts: the circular and linear paths as in Fig. 1(b). For convenience, the two half-circles are connected to each other, so are the linear paths. Both ends of the straight line in Fig. 1(b) are at the same place, though they are separated. Equivalently the light beams can be considered to traverse one circle and one straight line. Then the difference $\Delta t_L$ between the arrival times of $b_+$ and $b_-$, as seen from $S$, can be expressed as

$$\Delta t_L = \frac{2(l_c + l_s)\beta}{c(1-\beta^2)} \tag{16}$$

where $\beta = v/c$ and $l_c$ and $l_s$ are the lengths of the circular and linear, respectively, paths in $S$.

The velocities of $b_+$ and $b_-$ in the linear path can be found by using (4). In special relativity, the transformation coefficients in (4) are given by

$$a = \gamma^{-1}, \ \ b = \gamma, \ \ d = 1 \tag{17}$$

where $\gamma = (1-\beta^2)^{-1/2}$. One of the most important concerns in the test theory is to find the vector parameter $\boldsymbol{\varepsilon}$ which reflects physical reality. However, we do not deal with the matter of clock synchronization, and employ the standard synchronization. When the standard synchronization is used, the synchronization vector becomes $\boldsymbol{\varepsilon} = -\boldsymbol{\beta}$. Then $g$ and $\boldsymbol{\rho}$ are expressed as $g = \gamma$ and $\boldsymbol{\rho} = -\gamma\boldsymbol{\beta}$. With these coefficients and parameters, $\boldsymbol{T}_G^T \boldsymbol{T}_G = \boldsymbol{I}$. It is convenient to introduce a partitioned matrix which will be used in place of transformation matrices:

$$\boldsymbol{A} = \begin{bmatrix} A_{11} & \boldsymbol{A}_{12} \\ \boldsymbol{A}_{21} & \boldsymbol{A}_{22} \end{bmatrix} \tag{18}$$

where $A_{11}$ is a scalar quantity. Let $\boldsymbol{A} = \boldsymbol{T}_G$, and then $\boldsymbol{x}'$ is rewritten as

$$\boldsymbol{x}' = \boldsymbol{A}_{21}\tau + \boldsymbol{A}_{22}\boldsymbol{x}\,. \tag{19}$$

We can consider two methods in measuring the elapsed time when a light beam travels along the linear path. One is to measure it with respect to AT, the standard-synchronized time, along the path in the forward or reverse direction, and the other is to employ PT of the detector $O'$. Hereafter we use a subscript '$\circ$' in PT, say $\tau'_\circ$, to distinguish it from AT. It is well known that the differential time $d\tau'_\circ$ is related to $d\tau$ by

$$d\tau'_\circ = \frac{d\tau}{\gamma}\,. \tag{20}$$

The PT interval, which is measured at the same place, is regarded as absolute as its value is invariant in every inertial frame irrespective of $\boldsymbol{\varepsilon}$. However, the AT interval, which is the time difference between different places, depends on $\boldsymbol{\varepsilon}$. These facts imply that the PT would be appropriate for the measurement of the speeds.

The line segments $P_0P_1$ and $P_4P_5$ in Fig. 1(a) belong to the same inertial frame, which is different from that of $P_2P_3$ because the directions of the velocities are different though the speeds are equal. Suppose that they belong to the same inertial frame $S'$. The difference in direction is considered in Subsection 3.2. To derive the speed of light with respect to PT in $S'$, we calculate $d\boldsymbol{x}'$, which is written from (19) as

$$d\boldsymbol{x}' = d\tau(\boldsymbol{A}_{21} + \boldsymbol{A}_{22}\boldsymbol{c}_\tau) \tag{21}$$

where $\boldsymbol{c}_\tau = d\boldsymbol{x}/d\tau$. The velocity of a light beam with respect to $\tau'_\circ$ in $S'$ is expressed from (20) and (21) as

$$\frac{d\boldsymbol{x}'}{d\tau'_\circ} = \frac{d\boldsymbol{x}'}{d\tau}\frac{d\tau}{d\tau'_\circ} = \gamma(\boldsymbol{A}_{21} + \boldsymbol{A}_{22}\boldsymbol{c}_\tau)\,. \tag{22}$$

According to (5) and (6), $\boldsymbol{A}_{21}$ and $\boldsymbol{A}_{22}$ are independent of $\boldsymbol{\varepsilon}$. It should be noted that the velocity with respect to PT is independent of $\boldsymbol{\varepsilon}$. The squared norm of $d\boldsymbol{x}'$ is given by

$$\| d\boldsymbol{x}' \|^2 = d\tau^2(\boldsymbol{A}_{21}^T\boldsymbol{A}_{21} + 2\boldsymbol{A}_{21}^T\boldsymbol{A}_{22}\boldsymbol{c}_\tau + \boldsymbol{c}_\tau^T\boldsymbol{A}_{22}^T\boldsymbol{A}_{22}\boldsymbol{c}_\tau)\,. \tag{23}$$

Since $\boldsymbol{A}^T \boldsymbol{A} = \boldsymbol{I}$, it follows that

$$\boldsymbol{A}_{21}^T \boldsymbol{A}_{21} = 1 - A_{11}^2, \quad \boldsymbol{A}_{21}^T \boldsymbol{A}_{22} = -A_{11} \boldsymbol{A}_{12}, \quad \boldsymbol{A}_{22}^T \boldsymbol{A}_{22} = \boldsymbol{I} - \boldsymbol{A}_{12}^T \boldsymbol{A}_{12}. \tag{24}$$

From (23) and (24), we have

$$\| d\boldsymbol{x}' \|^2 = d\tau^2 [(1 - A_{11}^2) - 2 A_{11} \boldsymbol{A}_{12} \boldsymbol{c}_\tau + \boldsymbol{c}_\tau^T (\boldsymbol{I} - \boldsymbol{A}_{12}^T \boldsymbol{A}_{12}) \boldsymbol{c}_\tau]. \tag{25}$$

For a light signal, $\| d\boldsymbol{x} \| / dt = c$ and thus $\| \boldsymbol{c}_\tau \|^2 = -1$. Recall that $\| d\boldsymbol{x} \| > 0$. Substituting $\| \boldsymbol{c}_\tau \|^2 = -1$, $A_{11} = \gamma$, and $\boldsymbol{A}_{12} = -i\gamma \boldsymbol{\beta}^T$ into (25) and taking the square root of both sides yield

$$dl' = -\gamma \, d\tau (i + \boldsymbol{\beta}^T \boldsymbol{c}_\tau) \tag{26}$$

where $dl' = \| d\boldsymbol{x} \|$. Then the speed of light, $c' = dl' / dt'_\circ$, is obtained as

$$c' = ic \frac{dl'}{d\tau} \frac{d\tau}{d\tau'_\circ} = \gamma^2 c (1 - \boldsymbol{\beta}^T \hat{\boldsymbol{c}}) \tag{27}$$

where $t'_\circ = \tau'_\circ / ic$. Note that we have derived (27) in the standard synchronization. The propagation directions of $b_+$ and $b_-$ are parallel and antiparallel, respectively, to $\boldsymbol{v}$, as seen in $S$. The speeds of $b_\pm$ in $S'$ are given by $c'_\pm = \gamma^2 (c \mp v)$, respectively, which correspond to those derived from TCL in [7].

The speed of light in $S'$ is other than $c$, and becomes $c$ if measured with respect to $t'$ rather than $t'_\circ$. It can be easily seen by using (26) and $d\tau' = -i\gamma d\tau (i + \boldsymbol{\beta}^T \boldsymbol{c}_\tau)$, which results from $\tau' = A_{11} \tau + \boldsymbol{A}_{12} \boldsymbol{x}$, that $dl' / dt' = c$. If the speeds of $b_\pm$ were really $c$, there would be no time difference and thus no Sagnac interference. As can be seen from Fig. 1(a), it is obvious that the actual times taken during the travels of $b_\pm$ along the loop are the times measured by the clock of $O'$. The correct speeds must be measured with respect to PT, not AT.

### 3.2. Under the Consideration of Motion of Laboratory

In reality, as our Earth and Solar System move, the laboratory frame is not at rest and is different from the isotropic frame $S$. Though Earth rotates, we can consider that it belongs to an inertial frame during the very short test. We denote the laboratory frame by $S_i$ and an arbitrary inertial frame by $S_j$. The normalized velocity of $S_k$, $k = i, j$ is $\boldsymbol{\beta}_k = \boldsymbol{v}_k / c$ relative to $S$. The fiber loop that the light beams traverse is of arbitrary shape, as shown in Fig. 2. A curve can be approximated by many line segments. In Fig. 2, the loop is approximated by $n$ line segments, each of which is in linear motion at a constant speed of $v_j$ relative to $S$. As $n$ tends to infinity, the linearized shape approaches the original one. Generally the line segments belong to different frames since their directions of motion are different, though their speeds are the same.

Using (13) and (15), we have the $(1,1)$ -entry of $\boldsymbol{T}_G(\boldsymbol{\beta}_j, \boldsymbol{\beta}_i)$, which is written as

$$\boldsymbol{T}_G(\boldsymbol{\beta}_j, \boldsymbol{\beta}_i)|_{11} = \Gamma(\boldsymbol{\beta}_j, \boldsymbol{\beta}_i) \tag{28}$$

where $\boldsymbol{A}|_{mn}$, $m, n = 1, 2$, denotes the $(m, n)$ -entry of a partitioned matrix $\boldsymbol{A}$, i.e., $\boldsymbol{A}|_{mn} = \boldsymbol{A}_{mn}$. The entry $\boldsymbol{T}_G(\boldsymbol{\beta}_j, \boldsymbol{\beta}_i)|_{21}$ is given by [9]

$$\boldsymbol{T}_G(\boldsymbol{\beta}_j, \boldsymbol{\beta}_i)|_{21} = -i\Gamma(\boldsymbol{\beta}_j, \boldsymbol{\beta}_i)\boldsymbol{\beta}_{ij}\,. \tag{29}$$

Let $\boldsymbol{A} = \boldsymbol{T}_G(\boldsymbol{\beta}_j, \boldsymbol{\beta}_i)$. Given $A_{11}$, the entry $\boldsymbol{A}_{21}$ must be $-iA_{11}\boldsymbol{\beta}_{ij}$, as explained in the following. The differential coordinate vector of an object $O_i$ which is at rest in $S_i$ is represented as $d\boldsymbol{p}_{(i)} = [d\tau_{(i)}, \boldsymbol{0}^T]^T$. The differential vector in $S_j$ corresponding to the $d\boldsymbol{p}_{(i)}$ is calculated as $d\boldsymbol{p}_{(j)} = \boldsymbol{A}d\boldsymbol{p}_{(i)}$, which results in $d\boldsymbol{p}_{(j)} = d\tau_{(i)}[A_{11}, \boldsymbol{A}_{21}^T]^T$. The velocity in $S_j$ of $O_i$ is written as $\boldsymbol{\beta}_{ij} = id\boldsymbol{x}_{(j)}/d\tau_{(j)}$, where $d\tau_{(j)} = d\tau_{(i)}A_{11}$ and $d\boldsymbol{x}_{(j)} = d\tau_{(i)}\boldsymbol{A}_{21}$, which leads to $\boldsymbol{A}_{21} = -iA_{11}\boldsymbol{\beta}_{ij}$. One can also clearly see this relationship from the first column of (9) where $\boldsymbol{\beta}$ is the velocity in $S$ of an object $O'$ which is at rest in $S'$. The quantity $A_{11}$ corresponds to the time dilation factor and the PT of $O_i$ is related to $\tau_{(j)}$ by

$$\tau_{(i)\circ} = \frac{\tau_{(j)}}{A_{11}}\,. \tag{30}$$

Suppose that the standard synchronization is employed in $S_k$, $k = i, j$, so that $\boldsymbol{\varepsilon}_k = -\boldsymbol{\beta}_k$. For simplicity, we use $\gamma_k$ and $\gamma_{ji}$ instead of $(1-\beta_k^2)^{-1/2}$ and $\Gamma(\boldsymbol{\beta}_j, \boldsymbol{\beta}_i)$, respectively. Then $\boldsymbol{\beta}_{ij}$ is expressed, from (14) with subscripts $i$ and $j$ interchanged, as

$$\boldsymbol{\beta}_{ij} = \gamma_i \gamma_{ji}^{-1}[\gamma_j(\hat{\boldsymbol{\beta}}_j(\hat{\boldsymbol{\beta}}_j^T\boldsymbol{\beta}_i) - \boldsymbol{\beta}_j) + \boldsymbol{\beta}_i - \hat{\boldsymbol{\beta}}_j(\hat{\boldsymbol{\beta}}_j^T\boldsymbol{\beta}_i)]\,. \tag{31}$$

In the standard synchrony, $\boldsymbol{T}_G^T(\boldsymbol{\beta}_k) = \boldsymbol{T}_G^{-1}(\boldsymbol{\beta}_k)$, which leads to $\boldsymbol{T}_G^T(\boldsymbol{\beta}_j, \boldsymbol{\beta}_i) = \boldsymbol{T}_G^{-1}(\boldsymbol{\beta}_j, \boldsymbol{\beta}_i)$. Recall that $\boldsymbol{A} = \boldsymbol{T}_G(\boldsymbol{\beta}_j, \boldsymbol{\beta}_i)$. The differential spatial vector in $S_j$ is written from (11) as

$$d\boldsymbol{x}_{(j)} = \boldsymbol{A}_{21}d\tau_{(i)} + \boldsymbol{A}_{22}d\boldsymbol{x}_{(i)}\,. \tag{32}$$

Note that $d\tau_{(i)}^2 + \|d\boldsymbol{x}_{(i)}\|^2 = 0$ since the speeds of $b_\pm$ with respect to the AT, $t_{(i)}$, are $c$ in $S_i$. As $\boldsymbol{A}^T\boldsymbol{A} = \boldsymbol{I}$, (24) can be used. To use (24), it is necessary to discover $\boldsymbol{A}_{12}$. Interchanging $i$ and $j$ in (29) yields the relationship of $\boldsymbol{T}_G(\boldsymbol{\beta}_i, \boldsymbol{\beta}_j)|_{21} = -i\gamma_{ij}\boldsymbol{\beta}_{ji}$. Since $\boldsymbol{T}_G(\boldsymbol{\beta}_i, \boldsymbol{\beta}_j) = \boldsymbol{A}^T$, it follows that

$$\boldsymbol{A}_{12}^T = \boldsymbol{T}_G(\boldsymbol{\beta}_i, \boldsymbol{\beta}_j)|_{21} = -i\gamma_{ij}\boldsymbol{\beta}_{ji} \tag{33}$$

$$\gamma_{ji} = \gamma_{ij}\,. \tag{34}$$

From $\boldsymbol{A}^T\boldsymbol{A}=\boldsymbol{I}$, $A_{11}^2+\boldsymbol{A}_{21}^T\boldsymbol{A}_{21}=1$. Using this relationship, (28), and (29), we have $\gamma_{ji}^2(1-\beta_{ij}^2)=1$, which leads to

$$\gamma_{ji}=(1-\beta_{ij}^2)^{-1/2}. \tag{35}$$

From (34) and (35), $\beta_{ji}=\beta_{ij}$. Though the magnitudes of $\boldsymbol{\beta}_{ji}$ and $\boldsymbol{\beta}_{ij}$ are identical, generally $\boldsymbol{\beta}_{ij}\neq-\boldsymbol{\beta}_{ji}$. In case $\boldsymbol{\beta}_i$ and $\boldsymbol{\beta}_j$ are parallel or antiparallel, however, it is readily seen from (31) that $\boldsymbol{\beta}_{ij}=-\boldsymbol{\beta}_{ji}$. Using (33) and similarly following the computational procedure to find $c'$ above, we have

$$dl_j=-\gamma_{ji}d\tau_{(i)}(i+\boldsymbol{\beta}_{ji}^T\boldsymbol{c}_{i\tau}) \tag{36}$$

where $dl_j=\|d\boldsymbol{x}_{(j)}\|$ and $\boldsymbol{c}_{i\tau}=d\boldsymbol{x}_{(i)}/d\tau_{(i)}$.

The relationship (30) is valid even if $i$ and $j$ are interchanged. The speed of light with respect to $t_{(j)\circ}=\tau_{(j)\circ}/ic$ is given by

$$c_j=\frac{dl_j}{dt_{(j)\circ}}=\gamma_{ji}^2c(1-\boldsymbol{\beta}_{ji}^T\hat{\boldsymbol{c}}_i). \tag{37}$$

The speed of light is other than $c$. Though the one-way speed is not $c$, the two-way speed is $c$ regardless of the propagation direction. Equation (11) is a different representation of $\boldsymbol{p}_{(j)}$ which equals (4) with $\boldsymbol{p}'=\boldsymbol{p}_{(j)}$ and $\boldsymbol{\beta}=\boldsymbol{\beta}_j$. It is well known that the two-way speed in (4) is independent of $\boldsymbol{\varepsilon}$. The two-way speed of light is $c$ with respect to PT when the transformation coefficients are equal to (6). One can see similarities between (26) and (36) and between (27) and (37), which imply that actual physical quantities can be found through the standard synchronization that makes the laboratory frame $S_i$ appear isotropic as the preferred one $S$.

The *j*th segment of the loop in Fig. 2 is at rest in $S_j$. From (37), the speeds of $b_\pm$ in $S_j$ are written as $c_{j\pm}=\gamma_{ji}^2c(1-\boldsymbol{\beta}_{ji}^T\hat{\boldsymbol{c}}_{i\pm})$ where $\boldsymbol{c}_{i\pm}$ are the velocities of $b_\pm$ with respect to AT in $S_i$. The elapsed times during the travels of $b_\pm$ in $S_j$ are calculated as

$$dt_{(j)\circ\pm}=\frac{dl_j}{c_{j\pm}}=\frac{dl_j}{\gamma_{ji}^2c(1-\boldsymbol{\beta}_{ji}^T\hat{\boldsymbol{c}}_{i\pm})} \tag{38}$$

where $dt_{(j)\circ\pm}$ are the elapsed PTs when $b_\pm$ traverse the respective paths. Since $b_+$ and $b_-$ travel in the opposition directions, $\boldsymbol{c}_{i-}=-\boldsymbol{c}_{i+}$. The time difference in $S_j$ is given by

$$\Delta t_j=dt_{(j)\circ+}-dt_{(j)\circ-}=\frac{2dl_j\boldsymbol{\beta}_{ji}^T\hat{\boldsymbol{c}}_{i+}}{\gamma_{ji}^2c(1-(\boldsymbol{\beta}_{ji}^T\hat{\boldsymbol{c}}_{i+})^2)}. \tag{39}$$

As the direction of $\boldsymbol{c}_{i+}$ is identical with that of $\boldsymbol{\beta}_{ji}$, $\boldsymbol{\beta}_{ji}^T\hat{\boldsymbol{c}}_{i+} = \beta_{ji}$ and the denominator in (39) is reduced to $c$. The differential segments that compose the fiber loop move at the same speed of $v$, and thus all $\beta_{ji}$ are equal. Then

$$\Delta t_j = \frac{2dl_j\beta}{c} \tag{40}$$

where $\beta = \beta_{ji}$. As a result of it, the overall time difference $\Delta t_F$ observed at the detector is given by

$$\Delta t_F = \lim_{n\to\infty}\sum_{j=1}^{n}\Delta t_j = \frac{2l_F\beta}{c} \tag{41}$$

where $l_F = \lim_{n\to\infty}\sum_{j=1}^{n} dl_j$ is the rest length of the fiber loop. The time difference (41) corresponds to the experimental results [1, 2].

As $d\tau_{(j)\circ} = d\tau_{(i)}/\gamma$, the time difference observed in $S_i$ is given by $\Delta t_L = \gamma\Delta t_F$ where $\Delta t_L$ is the time difference with respect to AT. The length of the fiber loop as seen in $S_i$ is related to the rest length $l_F$ by $l_L = l_F/\gamma$ according to the well-known length contraction. It is seen by substituting $\Delta t_F = \Delta t_L/\gamma$ and $l_F = \gamma l_L$ into (41) that the resultant $\Delta t_L$ corresponds to (16), which represents the difference in AT. The detector in the experiment of the general Sagnac effect [1, 2] moves with the fiber loop and the effect of $\Delta t_F$ is measured.

So far we have analyzed the generalized Sagnac effect when $\boldsymbol{c}_{i+}$ has the same direction as $\boldsymbol{\beta}_{ji}$. Now it is time to examine a case where their directions are different. If the angle between $\boldsymbol{c}_{i+}$ and $\boldsymbol{\beta}_{ji}$ is $\theta$ so that $\boldsymbol{\beta}_{ji}^T\hat{\boldsymbol{c}}_{i+} = \beta_{ji}\cos\theta$, (39) can be written as $\Delta t_j = 2dl_j\beta_{ji}\cos\theta/c$ within a first-order approximation, and then (41) is given by $\Delta t_F = 2l_F\beta\cos\theta/c$. As far as the time difference is concerned, the effective length of the fiber loop reduces to $l_F\cos\theta$, as observed in the experiment [1], when the propagation direction of the light beam $b_+$ has an angle $\theta$ with respect to the direction of motion of the fiber loop.

## 4. Conclusion

The theoretical analysis, which corresponds to the experimental results, shows that the speeds of the counter-propagating light beams are different from $c$. In the global positioning system (GPS) also, the speed of light has been shown to be other than $c$ [11]. Accurate information on positioning can be obtained from GPS signals through the so called Sagnac correction. It is required because of the anisotropy of the light speed in the Earth frame. If it is really $c$, the Sagnac effect cannot occur. The generalized Sagnac effect shows the anisotropy in inertial frames as well as in rotating frames. One can make the light speed appear isotropic by applying the standard synchronization, for example, along the closed path $P_0$ through $P_5$ in Fig. 1(a). However, it is impossible to apply the clock synchronization to a closed path because the problem of time gap that multiple times, depending on paths, are defined at the same place [3, 4, 11] is caused. Moreover the standard synchrony cannot be set up for both the paths of $b_+$ and $b_-$ at the same time because their ATs are different. Even if it is carried out, the actual speed of light, which is measured with respect to PT, is different from $c$. The actual time elapsed during the travel is the time by the clock of the detector $O'$.

The anisotropy does not imply that the standard synchronization cannot be employed to discover physical quantities. On the contrary, the presupposition of the isotropy of inertial frames, but with its exact meaning, can allow us to readily approach the problems of physics, as shown in this paper. We approached the generalized Sagnac effect, viewing inertial frames as if they were isotropic. The standard synchrony leads to the property of $\boldsymbol{T}_G^T(\boldsymbol{\beta}_j, \boldsymbol{\beta}_i) = \boldsymbol{T}_G^{-1}(\boldsymbol{\beta}_j, \boldsymbol{\beta}_i)$ , which makes the mathematical manipulation easy. Moreover it can allow us to deal with relative motions between inertial frames without information on their velocities relative to $S$, as has usually been done. Clearly the speed of an object with respect to AT depends on the scheme of clock synchronization. When the useful standard synchrony is utilized, there is one important fact that PT is the correct time, and so the actual speed should be measured with respect to PT, not AT. With this fact, we have shown that physical quantities can be effectively found by regrading inertial frames as if they were isotropic in the standard synchronization.

## References

1. R. Wang, Y. Zheng, and A. Yao, Generalized Sagnac Effect, Phys. Rev. Lett., **93**, 143901 (2004).
2. R. Wang, Y. Zheng, A. Yao, and D. Langley, Modified Sagnac experiment for measuring travel-time difference between counter-propagating light beams in a uniformly moving fiber, Phys. Lett. A **312**, 7 (2003).
3. R. D. Klauber, Relativistic rotation: A comparison of theories, Found. Phys. **37**, 198 (2007).
4. G. Rizzi and M. L. Ruggiero, ed., *Relativity in Rotating Frames* (Kluwer Academic, Dordrecht, The Netherlands, 2004).
5. F. Hasselbach and M. Nicklaus, Sagnac experiment with electrons: Observation of the rotational phase shift of electron waves in vacuum, Phys. Rev. A **48**, 143 (1993).
6. A. Tartaglia and M. L. Ruggiero, Sagnac effect and pure geometry, Am. J. Phys. **83**, 427 (2015).
7. Y.-H. Choi, Coordinate transformation between rotating and inertial systems under the constant two-way speed of light, Eur. Phys. J. Plus **131**, 296 (2016).

8. R. Mansouri and R. U. Sexl, A test theory of special relativity: I. Simultaneity and clock synchronization, Gen. Relativ. Gravit. **8**, 497 (1977).
9. Y.-H. Choi, Systematic approach to the relativistic Doppler effect based on a test theory, Can. J. Phys. **94**, 1064 (2016).
10. M. Kretzschmar, Doppler spectroscopy on relativistic particle beams in the light of a test theory of special relativity, Z. Phys. A **342**, 463 (1992).
11. Y.-H. Choi, On constancy of the speed of light in the global positioning system, Phys. Essays **29** (3), 440–449 (2016).

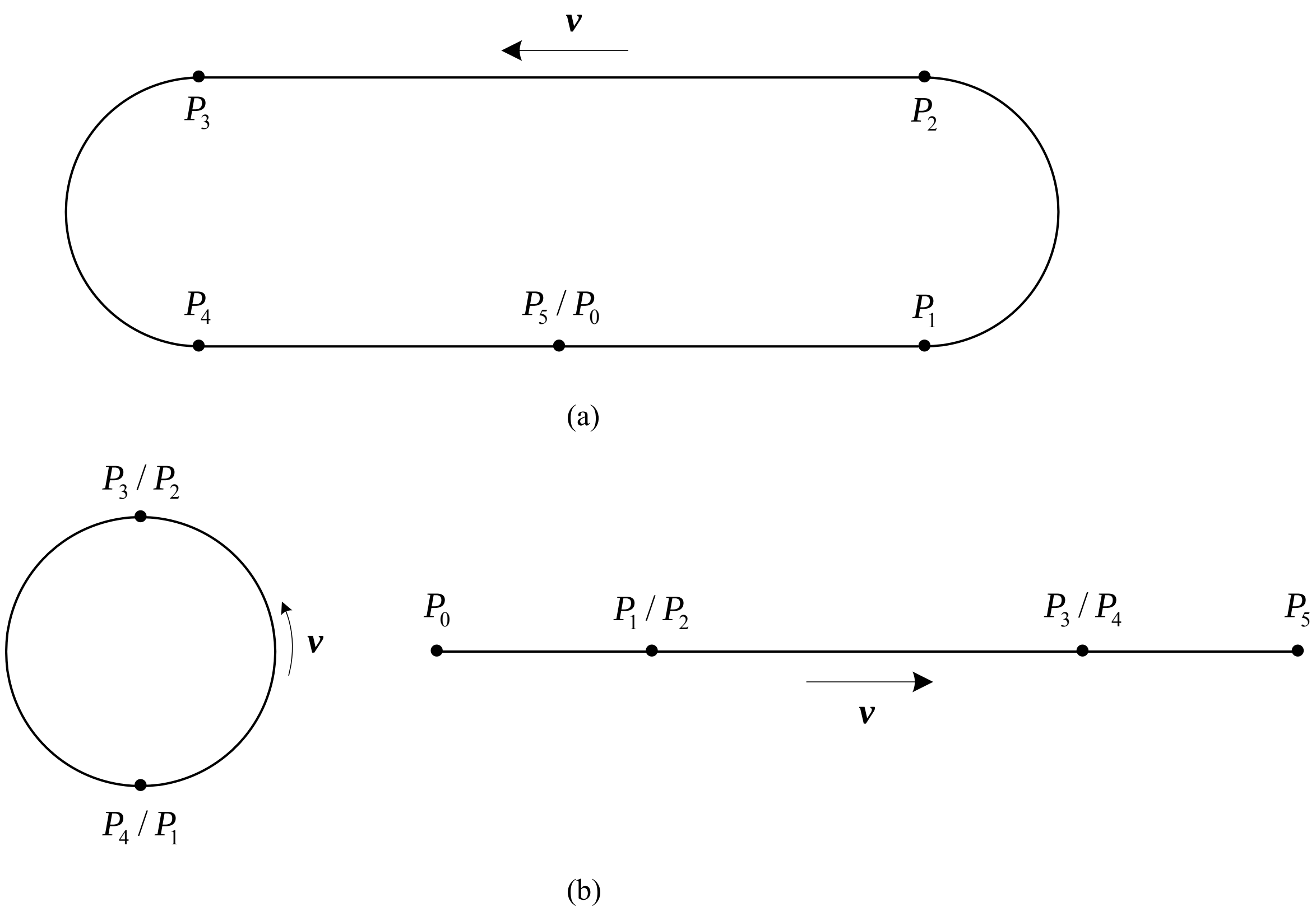

Fig. 1. (a) Optical fiber loop of simple shape. (b) Circular and linear paths.

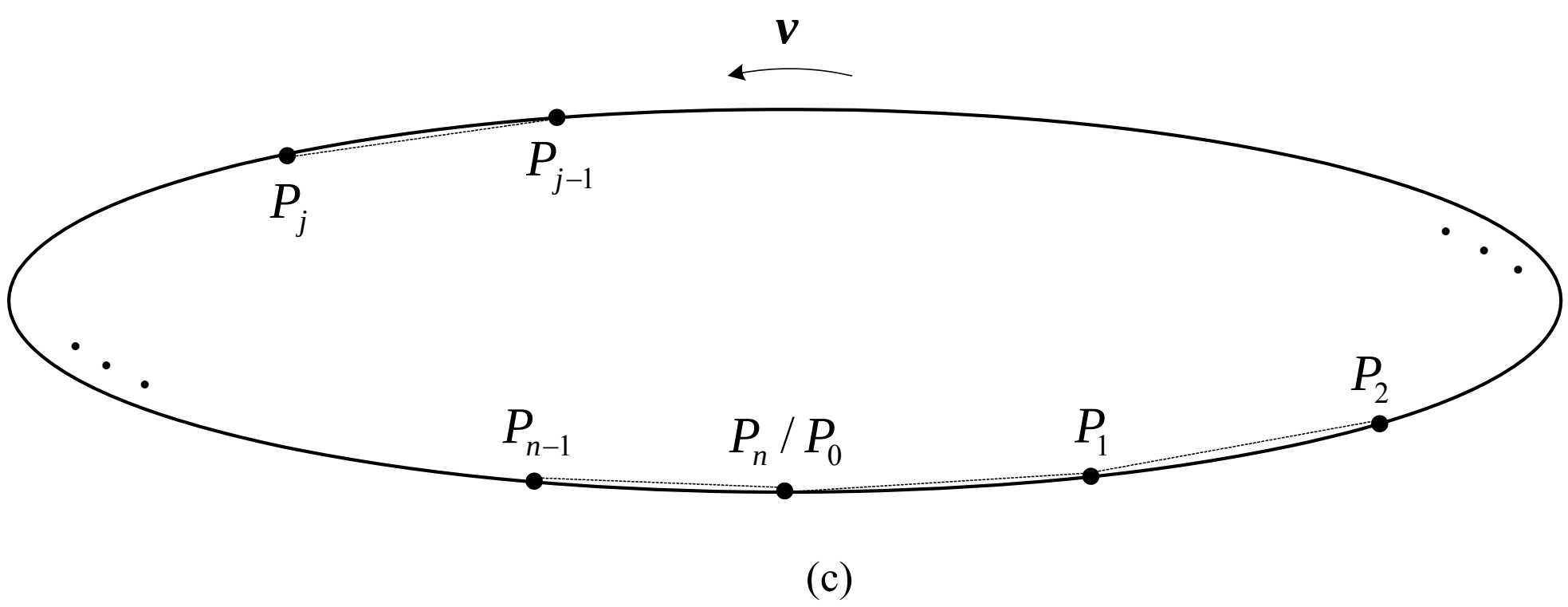

Fig. 2. Optical fiber loop of arbitrary shape.